# 23-bit Metaknowledge Template Towards Big Data Knowledge Discovery and Management

Nima Bari[*], Roman Vichr[**], Kamran Kowsari[†], Simon Berkovich[‡],
[*,‡,†] Department of Computer Science, George Washington University, Washington, DC 20052
[**] Exprentis, Inc Fairfax VA 22030
nbari@gwmail.gwu.edu[*], romanv@exprentis.com[†], berko@gwu.edu[‡]

*Abstract*— the global influence of Big Data is not only growing but seemingly endless. The trend is leaning towards knowledge that is attained easily and quickly from massive pools of Big Data. Today we are living in the technological world that Dr. Usama Fayyad and his distinguished research fellows discussed in the introductory explanations of Knowledge Discovery in Databases (KDD) [1] predicted nearly two decades ago. Indeed, they were precise in their outlook on Big Data analytics. In fact, the continued improvement of the interoperability of machine learning, statistics, database building and querying fused to create this increasingly popular science- Data Mining and Knowledge Discovery. The next generation computational theories are geared towards helping to extract insightful knowledge from even larger volumes of data at higher rates of speed. As the trend increases in popularity, the need for a highly adaptive solution for knowledge discovery will be necessary. In this research paper, we are introducing the investigation and development of 23 bit-questions for a Metaknowledge template for Big Data Processing and clustering purposes. This research aims to demonstrate the construction of this methodology and proves the validity and the beneficial utilization that brings Knowledge Discovery from Big Data.

## I. INTRODUCTION

Researchers and developers are currently aware of the difficulty in processing data as the datasets become larger and gain complexity. There are challenges associated with the influx of Big Data. For instance, according to [2], most businesses' metadata is arrayed across applications, databases, and "tribal knowledge" and managed by enterprises have no "formal metadata management system processes in place. Therefore, in theory there's a centralized mission but in practice there's a partition which separates Information Technology from Business. Hence, Metaknowledge drives the accuracy of reports, validates data transformations, ensures accuracy of calculations and enforces consistent definitions of business terms across multiple business users. Combined with the challenges Big Data proposes as it relates to time and memory complexities, new ideas for smart and insightful knowledge discovery are desperately needed. For this reason, Metaknowledge templates provide a visual representation of a process from its conceptualization to theory to implementation. The methodology can be applied to any dataset using tools employed for the goal of reaching an intended result. For the purpose of this presentation, we will refer to our sample dataset (Cinema Movies) which will navigate the processes outlined in the Metaknowledge template from preprocessing to the 23-bit Golay clustering. It is important to mention two important points: 1- The development of the 23 bits question template is not done arbitrarily but is done through a Metalearning process of Big Data Metaknowledge for feature selection and transformation. In addition to this, the process of formulating these 23 questions must fulfill the requirement of a) bias elimination and subjectivity and b) maximize probability of achieving the most accurate results; 2- the 23-bit template methodology can be applied and utilized for any domain knowledge, however, the specific 23-Bit questions are generated and applied for domain specificity to eliminate bias and maximize probability.

From another perspective, the Metalearning system plays a pragmatic role in optimizing the usage of Data Mining and Machine Learning tools. As a result, it turns their usage into a profitable business. Indeed, a Meta learning system builds upon the repetitive use of a predictive model over tasks to steer profits. Subject to data and statistics, this system has proved to provide major contribution to the successful adaptation of models to ever-changing needs. More specifically, a Meta Learning system deploys its unique pattern recognition function across tasks and ultimately serves as a control measure to the process of exploiting cumulative expertise [12].

## II. METAKNOWLEDGE AND METALEARNING

Metaknowledge is simply knowledge about knowledge. It allows data items to be examined inwardly for extraction of knowledge which can be used later. Furthermore, Metaknowledge can be utilized by an application or program allowing enhanced analysis of an initial data item for more informed results. From the Big Data point of view, the result is a data item representative of some formation, and the Metaknowledge is the information concerning the formation itself [3] (unclear). Early on, researchers in Artificial Intelligence mainly focused on developing one problem-solving paradigm that would be widely accepted and applicable. By the late 1960's, it was evident that such a paradigm is impossible to create because of the need to store large and domain specific knowledge [3]. For example, knowledge-based systems, which have been developed and captured amorphous data derived from tasks targeting specific areas like gaming, speech recognition, math, and chemistry, [4,5,6] etc. The large scale of task to collect the knowledge for these systems and the utilization and management of these large repositories has created a major research problem [3]. In addition, how to use and represent knowledge have both been problematic. Many encoding methods have been developed to direct and apply knowledge. Historically these tasks have been geared towards utilizing and applying knowledge for a particular domain [7, 8, 9, 10, 11].

Metalearning is by far an efficient tool to troubleshoot problems associated with the application of Data Mining (DM) and Machine Learning (ML) tools. Such troubleshooting effects are particularly witnessed in areas such as regression and classification. From a practical standpoint, the successful use of Data Mining and Machine Learning tools is inextricably linked to the proper selection of a predictive model(s) ; a model that fits to the model of application. It is very important to note that a lack of assistance in the process of model selection and combination or wrong Metafeatures can result in a wide range of obstacles to end-users [12]. These obstacles, including a user lack of expertise to choose the right model, tend to hinder both the cost-effectiveness and ease of the use of technology. In addition, end-users often face the limitations of models that operate on a trial-error basis. In order to tackle these barriers, a Metalearning system ensures both the automatic and systematic learning of user guidance. Unlike others systems, this learning system perfectly matches a specific task to a model or a combination of models [12].

From another perspective, the Metalearning system plays a pragmatic role in optimizing the usage of Data Mining and Machine Learning tools. More specifically, a Metalearning system deploys its unique pattern recognition function across tasks and ultimately serves as a control measure to the process of exploiting cumulative expertise through models [12].

### III. 23- BIT METAKNOWLEDGE VECTOR TEMPLATE FOR BIG DATA INSIGHTFUL PROCESSING AND CLUSTERING

Researchers and developers have been challenged in efficiently utilizing the hidden information from Metadata that comes with Big Data. Also called meaning of data, this hidden information can be localized through a simple observation of the desired clustering (posteriori), without prior knowledge of the applied clustering criteria. The computational Big data processing and clustering technique suggested here as a 23-bit template will allow the information retrieval of this valuable and insightful information[13].

The 23 meta-knowledge template processing technique to represent a record and to introduce its binary representation offers a unique added value such that it provides clusters of inter-related data objects quickly and in a linear timeframe. The database is scanned together by answering the 23 questions template (derived from Metaknowledge) in order to provide meaningful clusters of records. The 23-bit questions' Metaknowledge is inspired from the popular 20 questions game [14]. The 23 questions are structured so that each can be answered with either a yes (1) or no (0). In other words, given a template called T=q1, q2, q3…q23, q is a single question representing a bit along the 23-bit vector. The assertion is that a 23 bit vector template is considered sufficient to characterize any type of information. While putting in place 23-bit templates on all sorts of high level activities, this suggested technology is subject to distinct structuring of ontology interpretation; an unprecedented breakthrough that did not take place during the science of language[14].

The Metaknowledge template approach interestingly allows the possibility to categorize an object based upon inherent characteristics such as image geometry, semantic interrelationships, and so on. It is very important to recall that today's main form of research is written-based. As a result, other types of research have become too complex to conduct since they fail to successfully pass the process of generating meaningful clusters. When other research fails, the Golay Coding approach becomes very useful. The challenge that arises from an extension of the conventional searching methodology hinges upon developing constructive attributes that are in turn used to develop the Metaknowledge of 23-bit vector templates [13].

The selection of 23-bit templates is not random. This is an outcome of a unique structure of the world- a mathematical oddity associated with exclusive properties of the perfect Golay code [23, 12, 7]. Several possible adaptations of the code-word construction have been investigated in [15]. The experimental results in [22] show, that with 23 bit binary code-word; a possible 277 representations with Hamming Distance 0,1,2 can be found. Per [22] the algorithm complexity is at O(n) at FuzzyFind Dictionary (FFD) and O(1) to access FFD. In order to take advantage of this approach the 23 bit [1,0] representation 23 attributes (Metafeatures) are determined per 23-bit vector. That 23-buit vector carries the Metaknowledge descriptor of the record itself in a given domain. This representation leads to creation of Goley Coding Hashes to index record and Goley Transformation Tables to represent a record for access [22]. Therefore the discovery and derivation of 23-bit attribute vector is essential to characterize the record itself.

In the following section we will give examples and show how we were able to generate these 23 bit record of binary questions/answers representing Metaknowledge of the given domain based on the Metalearning process for Metafeature attribute selection and consequent transformation.

### IV. META-KNOWLEDGE DISCOVERY, ATTRIBUTE/FEATURE DERIVATION

During the determination of questions to classify the stream of data we attempted to establish a framework-like process to avoid the pitfalls of representing meta-knowledge through possible domain knowledge, which is tempting on a small subset of data. Instead we adapted machine learning and data mining framework to derive and discover meta-knowledge contained in raw data. This Metaknowledge will allow us to classify input streams of data using Golay code to represent record as derived binary 23 bit vector for the purpose of indexing and accessing it.

As we evaluate various algorithms for Metaknowledge characterization [4], [5], it allows for 23-bit binary attribute representation in a vector for processing through a Goley Code algorithm. The record's vector is representative of Metafeatures. The Metafeatures are corresponding to a best fit series of questions (23 questions representing 23 Metafeature attributes of Metaknowledge). The single Metafeature attribute representation corresponds to a single bit in Golay code's 23-bit vector. This vector is utilized for the creation of Goley Coding Hashes to index records and Goley Transformation Table to access them [22].

The classification Algorithms used are summarized in the table below for the purpose of discovery.

TABLE I. APPLIED ALGORITHMS FOR METAKNOWLEDGE FEATURE SLECTION AND REFINEMENT

| Num. | Metaknowledge and meta-feature discovery algorithms | |
|---|---|---|
| | *Algorithm* | *Benefits* |
| 1 | Scatter Plots | > A quick assessment if the "average" pattern is linear, curved, or random<br>> If the trend is positive or negative association<br>> Strength of relationship<br>Identification of group of outliers (x,y) |
| 2 | Statistical Relationship | > Statistical relationship in variation of possible values of X and X.<br>> Regression equation to describe the "best" line through the data and to predict Y based on X<br>> If linear relationship anticipated then describe the strength and direction |
| 3 | Q-Q Plots | > Plotted to verify if a "normal distribution" is a good model.<br>> The loess line is used to assess whether the assumption of linearity is appropriate. |
| 4 | Decision Tree | > A fast learning curve, easy to test model and effective way to find "terminal" nodes.<br>> Purposeful test of quasi model by over-fitting to evaluate by adding/removing attributes.<br>> Assistance in deriving threshold driven questions to qualify answer Yes/No [1, 0]. |
| 5 | DMRT | > Discriminative classifier making validation of threshold setting in questions.<br>> A wrong question would be indicated by no separation between class groups. |
| 6 | SVM | > Discriminative classifier to derive optimal hyper-plane on input data considering binary outcome. |
| 7 | GLM | A quick determination of attribute (questions) impact and contribution to the binary outcome. |

The sequence of algorithms applied provide guidance to determine Metaknowledge (for example, an indication of strength of relationships between attributes, ability to derive specific and concrete Metafeatures) [4], [5]. It also helps us make assertions about values (thresholds of boundaries) of feature questions. These questions are to be used as a foundation to represent the Metaknowledge in binary 23-bit word.

*A. Classification Margin*

As noted in [6], pp.34, several of the papers focus on applying algorithms to discover and extract Metaknowledge (sentence unclear). The [6] refers to a measurement of classification, which depends on number examples, algorithm error, and the parameter *k* as a user defined parameter determining size of the margin (perception of confidence). Such measures were helpful in comparing and quantifying differences among algorithms used and would be helpful in our next step to move to a semi-automated approach of Metaknowledge discovery translated into a series of questions for binary representation of a class based on Golay code.

The [6] presents a formula to determine what algorithm is performing well in Metaknowledge discovery. This can be used when multiple ranges of algorithms are used to determine the most suited Metafeatures for a given problem and domain.

The "formula" is to support the notion of qualification of "performing" well for an algorithm (while not applicable to all of them like scatter plots), which allows to set a margin for the set of algorithms utilized.

*B. Classification Framework*

We approach the problem first as a problem with a well-defined outcome (given the domain of movies, that is represented as Oscar Wins with Yes or No answers). This approach allows us to represent effectively a knowledge and corresponding record as binary Golay code word. That is translated into input in the Golay code as a sequence of 23 bit questions with binary [1, 0] answers. The attribute value is then set to 1 to represent "true" answer and set to "0" to represent negative answer. We make the assertion that by defining the problem as a problem with binary outcome (in this case represented by Oscar Wins) we make a discovery of Metaknowledge and Metaknowledge more standardized andrepeatable.. The outcome is then 23-bit binary vector representing record.

For our purpose we utilized extracted sample records from IMDB movie source [7], which contain a set of raw movie data points. These data points contained at the beginning unknown Metaknowledge, which had to be discovered (fragment). The framework of discovery is discussed in Section C. Through applied framework we attempted to increase information entropy of the record while minimizing the degree of freedom in given meta-feature attributes (which is done through the binary representation).

We also assert that the discovered Metaknowledge from a given data sample can be applied to all records in the domain. Applying Bayes algorithm assumption of inferring a "complete" knowledge from raw or Metafeatures knowledge to classify a full record accurately [9] we see the chance of getting result "A" given a positive test for "X", where X is not a zero.

$$P(A|B) = \frac{P(B|A)P(A)}{P(B)}.$$

The prior knowledge of movie sample records (while attractive on small subset of records) was represented as binary questions of what exact actor or director received an Oscar win. It initially appeared as the right classifier to use. However, in the consequent evaluation to generate Metafeatures by algorithm(s) (Table I.) these turned out to represent minimal information entropy about the data record and were rather classified as weak Metaknowledge features.

*1) Enhanced datasets ( a posterior Metaknowledge).*
Since the raw data did not contain Metafeatures for the purpose of deriving a posterior meta-knowledge we re-applied algorithms in Table I. We made this assertion with the purpose of deriving new additional features under the assumption of Shannon's distribution of information entropy [8]. Each meta-feature is represented as [1, 0] and consequently the distribution information is as "log of 2". Therefore all probability of distribution $p_i$ is positive or NULL, and consequently the SUM $(p_i.) = 1$ (formula below [8], [22]).

$$h(x) \sim k \sum p_i \log \frac{1}{p_i},$$

This principle justifies lowering the degree of freedom in meta-feature attribute, while increasing predictability of quasi-outcome. This allows us to apply Golay code as a means of classification and improve overall efficiency of given processing time.

C. Classification Result (all graphs/explanations)

Discovery of enhanced attributes - Enhanced datasets (a posterior meta-knowledge)

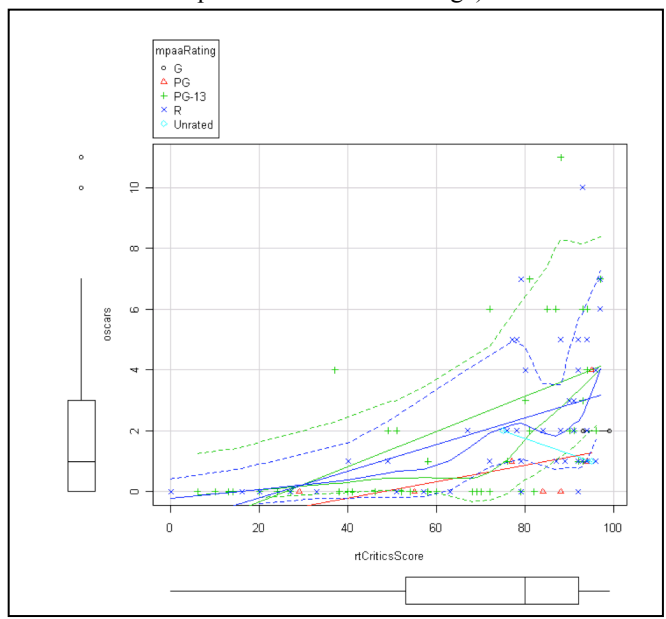

Figure 1.: Scatter Plot (correlation of Oscars vs rtCriticsScore by mpaaRating)

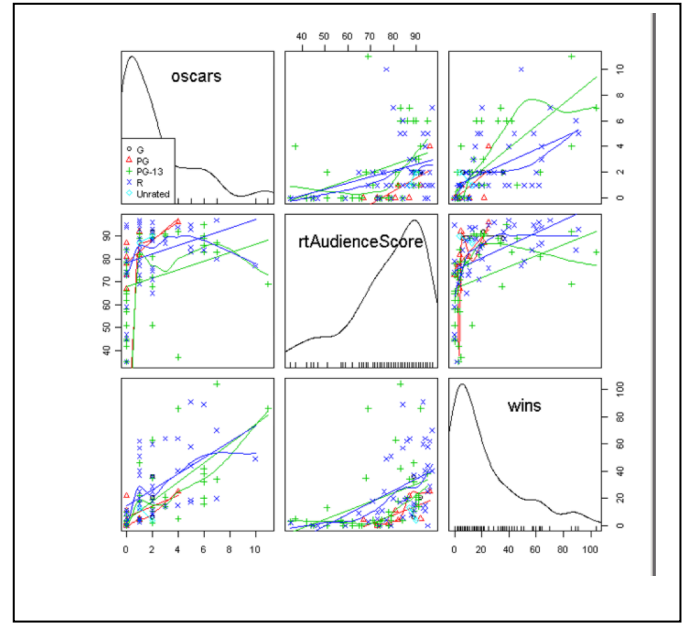

Figure 2.: Scatter Plot (correlation of OscarWins, rtAudienceScore, Wins)

The correlation graph Fig…. below shows a correlation graph (repetitive) between critic's Score and Audience Score grouped by Oscars attribute (where a number corresponds to the number of Oscars awards received). The area above the 80th percentile is the area of interest showing strong correlation between high CriticsScore, AudienceScore and Oscar wins.

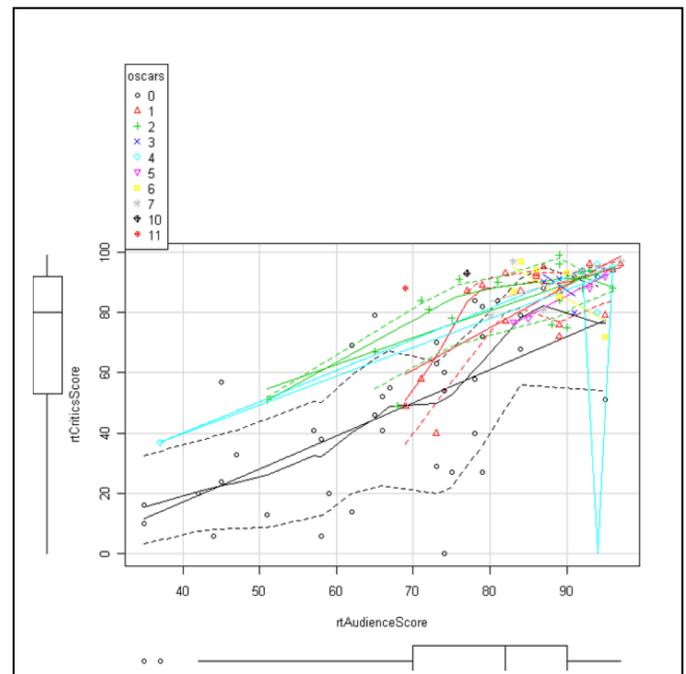

Figure 3.: A scatter plot of CriticsScore vs AudienceScore by Oscar wins.

Traditionally the distribution of pattern in records is assessed using histograms or cumulative distribution function (CDF). We approached it by using Q-Q (Quantile-Quantile) plots. The Figure 4 shows Q-Q plot of imdbRating shows correlation to outcome with rating above 6 and below 8.8 fitting linear model (wording).

Figure 4.: Q-Q Plot of imdbRating .

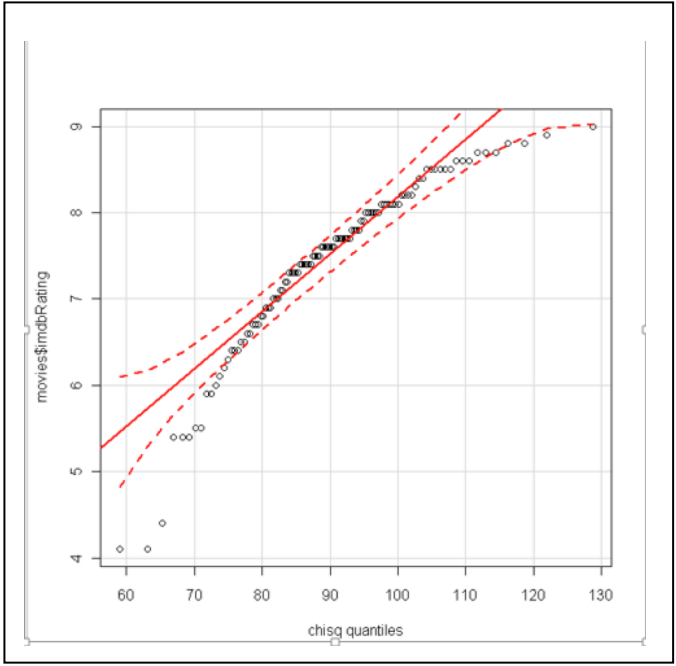

The Q-Q plot of imdbRating shows a range of 5.8 to 9 as the main range leading to outcome of tested Oscar award (as shown above Figure 3.) (sentence unclear). It shows imdbRating attribute as one to rely on in deriving the binary question. In the next step we subject this attribute to model testing (the case of overfitting is not that critical because we test feasibility and importance of meta-feature). We used this to derive a threshold value to tune a binary question to provide binary answer. Similarly we applied Q-Q plot to "movies.AudienceScore", which shown the range between 55 to 95 percentile mainstream with likelihood of classification to receive an Oscar award. Consequently, one can construct a binary answer question derive a Metafeature attribute classification, for example "Was the movie.AudienceScore above 55?. The answer can be Yes or No represented as [1,0] for the purpose of constructing a binary vector.

The Q-Q plot Figure 5 was used to confirm our assertion that it is correct to classify the as problem with binary outcome [1, 0]. The zero [0] is clearly defined and distinct outcome (fragment). Therefore the outcome of "1" is assigned to any movie that was awarded an Oscar in any movie category.

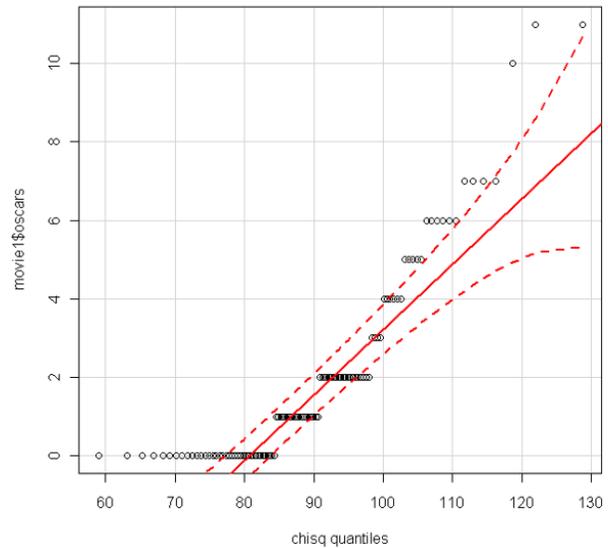

Figure 5.: Q-Q plot of Oscar wins

Based on the sequence of evaluations we determined, for example, the following attributes to be useful and benefiting our quasi model. It was criticsScore and imdbRating both being correlated to group defined as "Wins_GT4", which had a value "1" for number of wins > 4 and "0" for wins <0, i.e. representing a question with [Yes,No] answer. The "wins_GT4 "is an example of a derived attribute based on the Metaknowledge in raw data. The graph …… below shows the area in yellow, which is the area of strongest correlation between imdbRating and scriticsScore given "Wins_GT4" = 1".

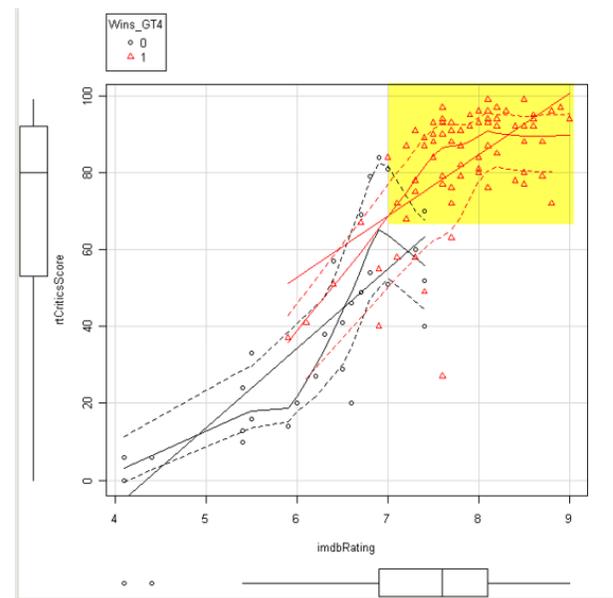

Figure 6.: Scatter plot of CriticsScore vs indbRating by "Wins_GT4".

The resulting Metafeature attribute derived was "What was the number of wins GT 4?", which could then we answered as Yes, No [1,0] for the purpose of Golay code representation.

We applied several variations of decision trees to test meta-feature attributes and to revalidate the thresholds used in discovery of these Metafeatures. For this purpose we focus on terminal nodes of a decision tree. This way we derive a rule for threshold of a new metaattribute. The figure below shows the "regression tree" for Model C which utilized several discovered meta-feature attributes like "Nominations_gt20", "Number_wins_GT15", or "imdb_rating>=7.5".

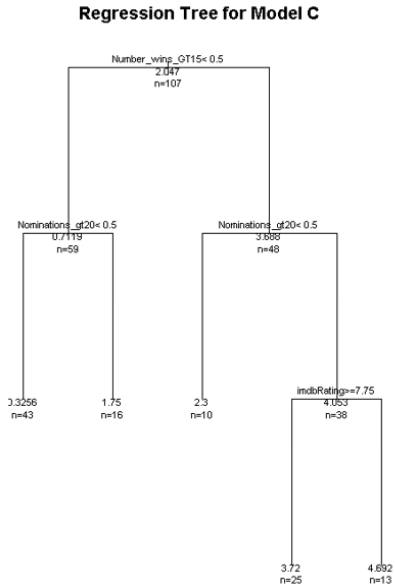

Figure 8.: Regression Tree based on Model C definition (incorporated newly discovered meta-knowledge based Metafeatures).

*1) Decision Trees*

We also applied a conditional tree model technique to the input set of attributes (questions) that were evaluated for the best fit and to determine terminal nodes.

Figure 7.: Conditional tree (model C) using discovered Metafeatures.

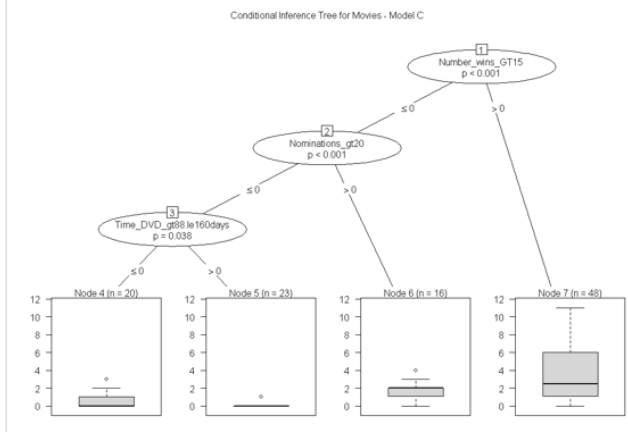

The Golay code work is represented as a set of 23 attributes (parameters) to be represented as Yes, No [1,0] and the "decision tree" algorithms well suited for determination of Metafeatures through deriving conditions of terminal nodes on classification of data. We made a claim that the issue of over-fitting into quasi model here does not pose a problem for classification.

*2) SVM (Support Vector Machine)*

The SVM method was effective in determining what initial set of features like 'Francis_Fisher' and 'David_O_Russell' and 'Box_office_sales_GT50_LE120mil' were determined as not suited to represent record. However the CriticsScore_GT40 as meta-feature was identified as high impact on outcome as well as Audience_*Score_GT60 and Nominations_gt20 and Wins_GT4*. The method was well suited to discover hidden meta-knowledge pattern and to confirm the selected Metafeatures.

*3) GLM (General Linear Model)*

This technique is used to quickly discover attribute and meta-feature is strongly impacting the outcome (wording).

```
The variable with significant impact:
CriticsScore_GT40    17.26266  2304.79562   0.007  0.9940
Audience_Score_GT60  -18.00375 2304.79615  -0.008  0.9938
Wins_GT4              2.35210     1.40571   1.673  0.0943 .
```

Figure 9.: GLM output of the most significant and discovered Metafeatures

The reason is to test easily and efficiently what meta-feature is to be evaluated more thoroughly (needs more explanation). In one run we subjected to GLM method test the whole set of 26 attributes that were subjects the of GLM method evaluation to confirm significant impact of attributes with [1,0] classification (i.e. low degree of freedom) on the intended outcome. The GLM method provided a quick assessment of the attributes derived from raw data as well as discovered meta-feature attributes and their impact on simulated outcome (Oscar wins in our considered dataset).

*D. Classification Conclusions and Summary*

Table II is a summary and reflects the results of the applied framework to the raw data. It enlists an example of meta-feature attributes.

TABLE II. FINAL SET OF QUESTIONS

| Attribute | Derivation of 23-bit attribute vector | | |
|---|---|---|---|
| | *Metafeature Question* | *Answer [Y, N]* | |
| Q1 | Any Oscar award? | Y | N |
| Q2 | Any Nominations? | Y | N |
| Q3 | Any nomination awards > 20 ? | Y | N |
| ….. | ……… | | |
| Q22 | Any wins > 4 | Y | N |
| Q23 | Is Audience Score > 55 | Y | N |

The derived set of Metaknowledge of the domains leads to selection of the best questions representative of the records (Table II.) and the given domain and leads to 23-bit binary attribute vector representation (Figure 8), which is fed into Goley Code based algorithm to generate Goley Coding Hashes to index records and Goley Transformation Tables to access the records.

Figure 8.: Goley code 23-bit attribute vector representation.

| Q1 | Q2 | Q3 | ... ........ | Q21 | Q22 | Q23 |
|----|----|----|--------------|-----|-----|-----|
| 1  | 0  | 1  | ... ........ | 0   | 0   | 1   |

## V. CONCLUSION AND FUTURE WORK

The established framework was instrumental to discover Metaknowledge and to derive Metafeatures for binary vector representation of attributes to process in Goley Code.

We assert that such framework and Metaknowledge discovery approaches combined with Goley Code processing can be reliably applied in general to Big Data problem (indexing, clustering, and access). The correct derivation of Metafeature attributes for 23-bit vector representation is critical to proper indexing, clustering and access. The Metafeature related conclusions can then be extrapolated to Big Data based on sampling theories [18, 20] in a domain specific "point of view". Such extrapolation of Metafeature characteristics to Big Data represented as a 23-bit binary vector in Golay code will lead to reliable classification regardless.

We make the assertion that for the purpose of qualifying algorithm (like Golay code in our case) the problem is best approached as a problem with binary [1, 0] outcome.

While some techniques, like decision trees, bring the notion of quasi-complete data separation issues, and thus questionablele validity of the model fit, it is not perceived as a negative impact in this case during the discovery of the Metafeatures to represent records (too many words, simplify). Simply, we desire to maximize (minimize?) over-fitting to avoid focusing on lesser impact Metafeatures. Therefore we are able to discover the driving meta-knowledge (driving patterns, factors and attributes) to represents the record in Golay code.

The Automation of evaluation performing algorithms detects "optimization and performance" in regards to bit classification attributes on the fly for a final set of meta-knowledge 23 attributes. Those are then used as input template of 23 questions to classify input stream of data using Golay code.

The framework (sequence of algorithms) to evaluate and characterize meta-knowledge appears stable and reliable, to produce reliable classifiers in the form of questions, each representing a single bit of Golay code.

The next step is to provide a theoretical background to the meta-knowledge characteristics discovery which minimizes the error in classification and show how Golay code can ensure stable and consistent classification despite possible errors (mischaracterization) in few bits of the word itself.

## *Acknowledgment*

We would like to sincerely thank Professor Simon Berkovich for his guidance and a special thanks to our friend and collaborator Dr. Roman Vichr.